\documentclass[aps,prd,amsmath,amssymb,footinbib]{revtex4-1}

\usepackage{graphicx}
\usepackage{natbib}

\usepackage{bm}
\usepackage{url}
\usepackage{textcase}
\usepackage{subcaption}
\usepackage{float}
\usepackage{amsthm}
\usepackage{braket}
\usepackage{color}
\usepackage[normalem]{ulem}
\usepackage{xcolor}

\newcommand{\be}{\begin{equation}}
\newcommand{\ee}{\end{equation}}

\begin{document}

\title{Spontaneous nonlinear scalarization of Kerr black holes }

\author{Daniela D. Doneva}
\email{daniela.doneva@uni-tuebingen.de}
\affiliation{Theoretical Astrophysics, Eberhard Karls University of T\"ubingen, T\"ubingen 72076, Germany}

\author{Lucas G. Collodel}
 \email{lucas.gardai-collodel@uni-tuebingen.de}
\affiliation{Theoretical Astrophysics, Eberhard Karls University of T\"ubingen, T\"ubingen 72076, Germany}

\author{Stoytcho S. Yazadjiev}
\email{yazad@phys.uni-sofia.bg}
\affiliation{Theoretical Astrophysics, Eberhard Karls University of T\"ubingen, T\"ubingen 72076, Germany}
\affiliation{Department of Theoretical Physics, Faculty of Physics, Sofia University, Sofia 1164, Bulgaria}
\affiliation{Institute of Mathematics and Informatics, 	Bulgarian Academy of Sciences, 	Acad. G. Bonchev St. 8, Sofia 1113, Bulgaria}

\begin{abstract}
As it became well known in the past years, Einstein-scalar-Gauss-Bonnet (EsGB) theories evade no-hair theorems and allow for scalarized compact objects including black holes (BH). The coupling function that defines the theory is the main character in the process and nature of scalarization. With the right choice, the theory becomes an extension of general relativity (GR) in the sense any solution to the GR field equations remains a solution in the EsGB theory, but it can destabilize if a certain threshold value of the spacetime curvature is exceeded. Thus BHs can spontaneously scalarized. The most studied driving mechanism to this phenomenon is a tachyonic instability due to an effective negative squared mass for the scalar field. However, even when the coupling is chosen such that this mass is zero, higher order terms with respect to the scalar field can lead to what is coined nonlinear scalarization. In this paper we investigate how Kerr BHs spontaneously scalarize by evolving the scalar field on a fixed background via solving the nonlinear Klein-Gordon equation. We consider two different coupling functions with higher order terms, one that yields a non-zero effective mass and another that does not. We sweep through the Kerr parameter space in its mass and spin and obtain the scalar charge by the end of the evolution when the field settles in an equilibrium stationary state. When there is no tachyonic instability present, there is no probe limit in which the BH scalarizes with zero charge, i.e. there is a gap between bald and hairy BHs and they only connect when the mass goes to zero together with the charge.

\end{abstract}

\maketitle

\section{INTRODUCTION}

Over the past few year, Einstein-scalar-Gauss-Bonnet (EsGB) theories have received increasing attention. The reason lies on three main facts. First, these carry a simple modification to general relativity (GR), by the inclusion of the Gauss-Bonnet term which is only made of quadratic invariants of the curvature and is coupled to a single scalar field, yielding an extra degree of freedom at the level of the theory. Second, because this term alone is a topological one, with the proper choice of the coupling the theory simply extends GR instead of strictly modifying it, i.e. all solutions to GR are also solutions of these theories and in principle they could coexist in nature together with non-GR ones.  This is in constrast to the Einstein-dilaton-Gauss-Bonnet (EdGB) \cite{PhysRevD.54.5049}, that differs from EsGB only by the choice of the coupling  but offers no trivial solution for the scalar field. Finally, because the extra term is quadratic in the curvatures, any observable deviations from GR only occur in the strong field regime, so it is automatically in agreement with all the tests performed in the weak field. 

Given that these theories' new traits emerge only in the strong gravity domain, investigating black hole (BH) solutions is at the core of the community's interests. The first scalarized BH solutions to appear in the literature are static and spherically symmetric configurations that suffer from a tachyonic instability giving rise to a nontrivial scalar fields \cite{PhysRevLett.120.131103,PhysRevLett.120.131104}. Assuming a positive defined coupling function, this is always possible as long as its second derivative is greater than zero at the equilibrium point, i.e. where the coupling and its first derivative with respect to the scalar field vanish (which grants GR as part of the solution set). Rotating equilibrium solutions were thereafter also found for the same couplings \cite{PhysRevLett.123.011101,Collodel_2020}. As it was shown, with increasing spin the domain of existence of hairy BHs gets narrower as it tends to quench scalarization. The reason is that if the spin parameter is large enough, the Gauss-Bonnet term develops a series of negative local minima in the vicinity of the event horizon where the effective squared mass of the scalar field becomes positive, thus cancelling the tachyonic instability. In the same line of reasoning, if the BH spins fast enough and the second derivative of the coupling is negative instead of positive, the same instability arises triggered not only by the high curvature but also crucially by the spin. The first results to show that scalarzation does indeed occur for $a/M>0.5$ appeared in \cite{PhysRevLett.125.231101} where the authors expanded the angular dependence onto a basis set of spherical harmonics and performed a $1+1$ evolution on the perturbed equations. The same results were shortly later obtained by a $2+1$ evolution \cite{PhysRevD.102.104027,Doneva:2020kfv}, also at the linear level, and the threshold confirmed. Stationary solutions of spin scalarized BHs were reported in \cite{PhysRevLett.126.011104,PhysRevLett.126.011103}, as in the static and low rotation limit only GR BHs exist. Loss of scalar hair is also possible in a process called dynamical descalarization involving BH binary systems, where the remnant of hairy BH mergers might be bald  \cite{PhysRevLett.127.031101,East:2021bqk,Elley:2022ept,Doneva:2022byd} (see also \cite{Elley:2022ept} for spin-induced scalarization). It was demonstrated also that scalarized black holes can form as a results of a realistic astrophysical process such as stellar core-collapse \cite{Kuan:2021lol}. Beyond BHs, other compact object have been studied in these theories. Among them, scalarized neutron stars \cite{PhysRevLett.120.131104,Doneva_2018}, wormholes without exotic matter \cite{PhysRevD.101.024033} (as the effective energy momentum tensor already violates the weak equivalence principle), and particle-like solitonic configurations \cite{Kleihaus:2019rbg,PhysRevD.102.024070}, in which the scalar field diverges at the origin but all metric functions and curvature invariants remain finite.

Depending on the form of the coupling function, other form of instability can be accounted for, leading to a \emph{nonlinear} scalarization. In this case, there is no tachyonic instability and only a strong enough perturbation can give rise to a nontrivial scalar field around the black hole. This is always the case if the second derivative of the coupling function trivializes at the equilibrium point where the coupling and its first derivative with respect to the scalar field vanish. Thus, the GR black holes are always linearly stable. Backreacting BH solutions to the static EsGB field equations in this particular case were reported in \cite{PhysRevD.105.L041502} and some of the branches of solutions were proven to be stable \cite{Blazquez-Salcedo:2022omw}. Similar solutions were found also for charged black holes in Einstein-Maxwell-scalar gravity \cite{LuisBlazquez-Salcedo:2020rqp,Blazquez-Salcedo:2020nhs}.

The aim of the present paper is to advance the studies in  \cite{PhysRevD.105.L041502} to the case of rotating black holes. Solving the stationary field equations in Gauss-Bonnet gravity, and as a matter of fact any other quadratic theory of gravity, can be a formidable task. An alternative way to proceed is the following. Stationary solutions to the scalar field on a GR background can be found at the end state of evolution of the nonlinear Klein-Gordon equation, i.e. accounting for all contributions from the coupling function \cite{PhysRevD.103.064024}. As opposed to the linear level where one simply obtains the threshold for scalarization in the decoupling limit, with this approach one can \emph{approximately} access the ``equilibrium'' scalar field profile and obtain the corresponding scalar field charges. It is important to emphasize that these solutions are found on a fixed background and are therefore not backreacting. Hence, they do not need to satisfy regularity conditions that could eventually arise in the full set of equations. This means that the parameter space of the exact solutions could be narrower than the one obtained with the above described methodology. Similarly, the scalar field profile as well as the asymptotic charges are accurate enough only for weak scalar field and they could have a stronger deviation for large scalar fields.

In the present work, admitting two types of theories that undergo nonlinear scalarization, we perform a $2+1$ evolution of the scalar field on a Kerr background. The parameter spaces of the saturated solutions are presented in terms of the characteristic normalized spin. In Sec. \ref{sec:theory} we present the generals of EsGB theories and scalarization. The equation of motion in the chosen coordinate system is introduced in Sec. \ref{sec:background}. A brief overview of the numerical techniques we employ to solve for the system is given in Sec. \ref{sec:ns} and the results in Sec. \ref{sec:results}. We conclude in Sec. \ref{sec:conclusions}. Throughout this paper we adopt geometrical units such that $c=G=1$.

\section{Theory}
\label{sec:theory}
The action for the Einstein-scalar-Gauss-Bonnet gravity, in the absence of matter fields, reads
\be
\label{eq:action}
S=\frac{1}{16\pi}\int d^4x\sqrt{-g}\left[R-2\nabla_\mu\psi\nabla^\mu\psi+\lambda^2f(\psi)\mathcal{R}_{GB}^2\right], 
\ee
where $g$ is the metric determinant, $\psi$ is a scalar field which provides the extra degree of freedom to gravity and is coupled via the function $f(\psi)$ to the Gauss-Bonnet invariant $\mathcal{R}_{GB}^2\equiv R^{\mu\nu\sigma\rho}R_{\mu\nu\sigma\rho}-4R^{\mu\nu}R_{\mu\nu}+R^2$. In this last expression, $R_{\mu\nu\sigma\rho}$, $R_{\mu\nu}$ and $R$ are the Riemmann tensor, the Ricci tensor and curvature scalar, respectively. The equations of motion are obtained by varying this action with respect to the inverse of the metric, $g^{\mu\nu}$ and the scalar field $\psi$, leading to
\be
\label{eq:feq}
R_{\mu\nu}-\frac{1}{2}Rg_{\mu\nu}+\Gamma_{\mu\nu}=2\nabla_\mu\psi\nabla_\nu\psi-g_{\mu\nu}\nabla_\sigma\psi\nabla^\sigma\psi,
\ee
\be
\label{eq:kge}
\Box\psi=-\frac{\lambda^2}{4}f'(\psi)\mathcal{R}_{GB}^2,
\ee
where $\Box$ is the d'Alembert operator, the prime denotes a derivative with respect to the scalar field and we defined 
\be
\label{eq:gamma}
\Gamma_{\mu\nu}\equiv-2R\nabla_{(\mu}\Psi_{\nu)}-4\nabla_\sigma\Psi^\sigma\left(R_{\mu\nu}-\frac{1}{2}Rg_{\mu\nu}\right)+4R_{\mu\sigma}\nabla^\sigma\Psi_\nu
+4R_{\nu\sigma}\nabla^\sigma\Psi_\mu-4g_{\mu\nu}R^{\sigma\rho}\nabla_\sigma\Psi_\rho+4R^\sigma{}_{\mu\rho\nu}\nabla^\rho\Psi_\sigma,
\ee
with $\Psi_\mu\equiv\lambda^2f'(\psi)\nabla_\mu\psi$. 

The theory is only defined of course after fixing the coupling function $f(\psi)$ and its parameters. This choice is absolutely determinant for the nature of the theory and the existence conditions for the scalar field. For instance, admitting stationarity and asymptotic flatness, if $f$ is constant, one recovers general relativity and the scalar field is trivial everywhere. On the other hand, if $f\propto\psi$, the scalar field can only trivialize at zero curvature. We are interested in an extended theory that accommodates general relativity as a particular case. This is possible as long as $f(\psi_0)=0 \land f'(\psi_0)=0$ (C1, standing for condition 1) for some value of $\psi_0$ which we choose to be zero, without loss of generality. Hence, Kerr black holes are solutions to the equations (\ref{eq:feq})-(\ref{eq:kge}) and may be used as a background to evolve the scalar field in the nonstationary regime in order to assess spontaneous scalarization processes. 

Tachyonic instabilities arise if $f''(0)>0$ (C2), leading to the growth of scalar hair in strong gravity regimes, i.e. at vicinity of an event horizon, sourced by Gauss-Bonnet term. The tachyonic character of this instability reflects on the fact that from eq. (\ref{eq:kge}) $\mu^2\equiv-\frac{\lambda^2}{4}f''(\psi)\mathcal{R}_{GB}^2$ is an effective mass squared for the scalar field, which is then negative. This process was named \emph{curvature-induced scalarization} \cite{PhysRevLett.120.131103,PhysRevLett.120.131104}. Such phenomenon also occurs for rotating black holes, but it is quenched for high spin parameters as $\mathcal{R}_{GB}^2$ becomes negative in some domains and the squared mass becomes positive \cite{PhysRevLett.123.011101,Collodel_2020}. Conversely, if C2 is dropped and instead $f''(0)<0$ (C3) is adopted, the instability only kicks in for spin parameters $j\equiv a/M\geq 0.5$, the so called \emph{spin-induced scalarization} \cite{PhysRevLett.125.231101,PhysRevD.102.104027,Doneva:2020kfv} for which stationary solutions have also been found \cite{PhysRevLett.126.011104,PhysRevLett.126.011103}.

An entirely different scalarization mechanism can take place if neither C2 or C3 are satistied but $f''(0)=0$ (C4), corresponding to $\mu^2=0$. Clearly, there is no tachyonic instability anymore, but if the coupling contains higher order terms a \emph{nonlinear scalarization} occurs \cite{PhysRevD.105.L041502}. The essence of this effect is that Kerr black hole becomes unstable only if large enough initial perturbation is imposed.  As long as C1 is kept, this process is still spontaneous. Interestingly, different instabilities can coexist in one single system with a coupling function that satisfies $\mu^2<0$ but also contains higher order terms. It can then feature both tachyonic and nonlinear instabilities \cite{PhysRevD.105.L041502}.

In this work, two different couplings are considered:

\be
\label{eq:couplings}
f_1=\frac{1}{2\beta}\left(1-e^{-\beta\psi^2\left(1+\kappa\psi^2\right)}\right), \qquad f_2=\frac{1}{4\kappa}\left(1-e^{-\kappa\psi^4}\right),
\ee
where $\beta$ and $\kappa$ are dimensionless parameters and $f_1$ satisfies $C1$ and $C2$ while $f_2$ satisfies $C1$ and $C4$. Hence, the scalarization is spontaneous in both cases but while being both of tachyonic and nonlinear nature in $f_1$, it is exclusively nonlinear in $f_2$. We recall that these couplings have already been considered in the investigation of spontaneous scalarization of Schwarzschild BHs and that exact solutions have also been found, featuring a stable and an unstable branch \cite{PhysRevD.105.L041502,Blazquez-Salcedo:2022omw}. Here, we extend this work to the rotating case, by assessing the nonlinear evolution of the scalar field around Kerr BHs, i.e. in the decoupling limit. Even though this approach does not take into account the scalar field backreaction onto the metric, it turns out to give quite accurate results of the equilibrium scalar field profile and dynamics if the scalar field is weak enough \cite{PhysRevD.103.064024,Kuan:2021lol,Benkel:2016rlz,Ripley:2019aqj,Ripley:2020vpk}.

\section{Scalarization on a fixed background}
\label{sec:background}

In order to study the development of scalar hair in the theories of eq. (\ref{eq:action}) together with eq. (\ref{eq:couplings}), we performe a $2+1$ evolution of the Klein-Gordon eq. (\ref{eq:kge}) on a fixed background characterized by the Kerr metric. In Boyer-Lindquist (BL) coordinates it reads
\be
\label{eq:metric}
ds^2=-\frac{\Delta - a^2\sin^2\theta}{\Sigma}dt^2-\frac{4Mar\sin^2\theta}{\Sigma}dtd\phi+\frac{\left(r^2+a^2\right)^2-\Delta a^2\sin^2\theta}{\Sigma}\sin^2\theta d\phi^2+\frac{\Sigma}{\Delta}dr^2+\Sigma d\theta^2,
\ee
where $\Delta=r^2-2Mr+a^2$ and $\Sigma=r^2+a^2\cos^2\theta$. As usual, $M$ is the black hole's mass while the angular momentum is given by $J=aM$. The Gauss-Bonnet term for a Kerr black hole reduces to its Kretschmann scalar and is given by
\be
\label{eq:kerrGB}
\mathcal{R}_{GB}^2=\frac{48M^2}{\Sigma^6}\left(r^2-a^2\cos^2\theta\right)\left(r^4-14a^2r^2\cos^2\theta+a^4\cos^4\theta\right).
\ee

Instead of working directly with the coordinates in eq. (\ref{eq:metric}) it is convenient to keep its time and polar coordinates $\{t,\theta\}$ and adopt the Kerr-Schild angle $\varphi$ and tortoise radial coordinate $x$, with transformations
\be
\label{eq:coordtrans}
d\varphi=d\phi+\frac{a}{\Delta}dr, \qquad dx=\frac{r^2+a^2}{\Delta}dr.
\ee

The equation of motion for the scalar field in this new set of coordinates $\{t,x,\theta,\varphi\}$ spells out as 

\begin{align}
\label{eq:kge_cord}
&\left[(a^2+r^2)^2-\Delta a^2\sin^2\theta\right]\partial_{tt}\psi + 4aMr\partial_{t\varphi}\psi - (a^2+r^2)^2\partial_{xx}\psi - 2a\left(a^2+r^2\right)\partial_{x\varphi}\psi -  2\Delta r\partial_{x}\psi \nonumber\\
&- \Delta\left(\partial_{\theta\theta}\psi+\cot\theta\partial_{\theta}\psi+\csc^2\theta\partial_{\varphi\varphi}\psi\right)=\lambda^2\frac{12M^2}{\Sigma^6}\left(r^2-a^2\cos^2\theta\right)\left(r^4-14a^2r^2\cos^2\theta+a^4\cos^4\theta\right)f'(\psi),
\end{align}
one hyperbolic PDE in $3+1$ dimensions. Please note that $\partial_x\equiv\partial/\partial x$, $\partial_{tt}\equiv\partial^2/\partial t^2$ and similarly for the other coordinate derivatives. The axisymmetry of the background allows for configurations with harmonic dependence on the azymuthal angle, e.g. $\psi(t,x,\theta,\varphi)=\psi(t,x,\theta)e^{im\varphi}$.  However, the above equations is not linear for the couplings under investigation. We circumvent the difficulties this causes by simply evolving axisymmetric initial configurations, e.g. $m=0$. Such a choice affects the excited quasi-normal modes in the stable cases and the characteristic times but not the outcome of the scalarization process.

\section{Numerical Setup}
\label{sec:ns}

After dismissing the azymuthal dependence, the scalar field depends on two space component and on time. We solve the resulting equation via the method of lines. We adopt a finite differences method of second order to discretize the equation of motion in a two dimensional grid of $N=N_x\times N_\theta$ internal points. This transforms the problem to one of solving $N$ ordinary differential equations, which are then evolved in time with a Runge-Kutta method of third order. We have performed convergence tests with higher order discretization and integration methods to be certain this suffices. 

The spatial domain is such that in the polar direction $\theta\in[0,\pi]$ with usually $N_\theta=60$ internal points. However, in the radial direction the range and number of internal grid points are dependent on the background parameters $(M/\lambda, a/\lambda)$ which affect the instability times. In BL coordinates, let $r_H$ be the horizon radius. Starting from  $r_0\sim1.0001r_H$, the grid extends to a point in the range $r_\infty\in[500M,3000M]$ with $Nx\in[4000,24000]$ internal points.

The initial condition is always set to be a Gaussian pulse of width $\sigma\sim1.0M$ centered about $\mu\sim8.3M$. As for the boundary conditions, we have Neumann on the poles $\partial_\theta\psi|_{\theta=0,\pi}=0$, ingoing waves at the horizon and outgoing waves at infinity, also known as Sommerfeld boundary condition. A well-known issue with this condition at infinity is the spurious artificial reflecting signal. Our approach to this problem is to simply choose $r_\infty$ to be sufficiently large so that the returning wave never reaches the region of interest. Hence, the larger the instability times, the farther away $r_\infty$ must be.

In the perturbative linear case studied in \cite{PhysRevD.102.104027,Doneva:2020kfv} for instance, the integration is performed either until the scalar field trivializes indicating the stability of the background solution, or until it blows up showing the scalarization threshold. The nonlinear system at hand has the different property that the scalar field does not diverge when the background is unstable. Instead, it evolves to approach the stationary configuration (if it exists) that could be found by solving the system of equations (\ref{eq:feq}) and ({\ref{eq:kge}) when they are independent of time. This way, unlike in the linear case, we can extract a charge $D$ asymptotically from the final state as $\psi(t\sim\infty,r\sim\infty)\sim D/r$. Note that, for solely nonlinear scalarization, there is no probe limit corresponding to a stationary Kerr black hole with infinitesimal scalar hair of zero charge, meaning there is a gap between bald and hairy solutions. Naturally, because we are not backreacting, there are deviations from the true solutions but the general qualitative behavior is preserved. Furthermore, if the scalar field is small enough, its contribution to the geometry is negligible and the approximations are realistic enough \cite{PhysRevD.103.064024,Kuan:2021lol}. In any case, we perform the evolution for as long as it is required for stationarity to be achieved. 

\section{Results}
\label{sec:results}

The equation of motion given by (\ref{eq:kge_cord}) is solved with the initial and boundary conditions as stated above. Nevertheless, the dimensionless parameters that define the theory and background must be set in each case. These form the set $p=\{\beta, \kappa, M/\lambda, j\}$, where we define the normalized spin parameter $j\equiv a/M^2$. We fix the potential parameters as $\beta=24.0$, $\kappa=48.0$ for $f(\psi)=f_1(\psi)$ and $\kappa=\{1000.0, 10000.0\}$ for $f(\psi)=f_2(\psi)$, and sweep through the valid ranges of $M/\lambda$ and $j$ until scalarization stops. These choices are not arbitrary. The smaller the values of $\beta$ and $\kappa$ are, the larger should be the deviations from Kerr for the exact stationary scalarized solutions \cite{PhysRevD.105.L041502}. The adopted decoupling limit approximation, though, is only valid for weak scalar fields. Thus, in order that our approximations are still meaningful, we must restrict ourselves to parameter values that are not too small.

First, let us focus on the static solutions in order to demonstrate the domain of existence of nonlinearly scalarized black holes in the nonrotating limit. This also helps us demonstrate that that for the chosen values of the parameters the decoupling limit approximation is indeed both qualitatively and quantitatively accurate enough. The differences between sequences of solutions obtained with the evolving scalar field on a fixed Schwarzschild background and the static backreacting (exact) case are displayed in the charge vs. mass diagrams of Fig. \ref{Fig:0}. In the left panel we show the results for $f_1$ and on the right for $f_2$. In both cases there is a nonunique region where for a certain value of $M/\lambda$ two solutions of different charge $D/\lambda$ can be found by solving the elliptic equations. For such masses the Schwarzschild black hole is linearly stable. However, the lower branch is not accessible by evolving the scalar field, i.e. it is unstable. Moreover, the probe limit ($D/\lambda=0$, $M/\lambda\neq 0$) which appears on the left panel (but not on the right) occurs precisely on this unstable branch. 

Note that the domain on the right panel is considerably smaller than that on the left and that the increase in $\kappa$ further shrinks the region of existence. In addition, the values of $\kappa$ for $f_2$ are orders of magnitude higher than those chosen for $f_1$. The full spectrum of solutions for a wide range of $\beta$ and $\kappa$ can be found in \cite{PhysRevD.105.L041502} and the domain of existence of solutions is controlled by the coupling parameters. As we commented, though, the choice of the coupling parameters in the present paper is not arbitrary. Instead, $\beta$ and $\kappa$ are adjusted in such a way that as seen on the figure, the differences between the ``exact'' numerical solutions and the results from time evolution in the decoupling limit are small enough (apart from the low mass region, the difference is roughly $<10\%$ in the value of the scalar charge).

\begin{figure}[!ht]
\centering
\includegraphics[width=0.48\linewidth]{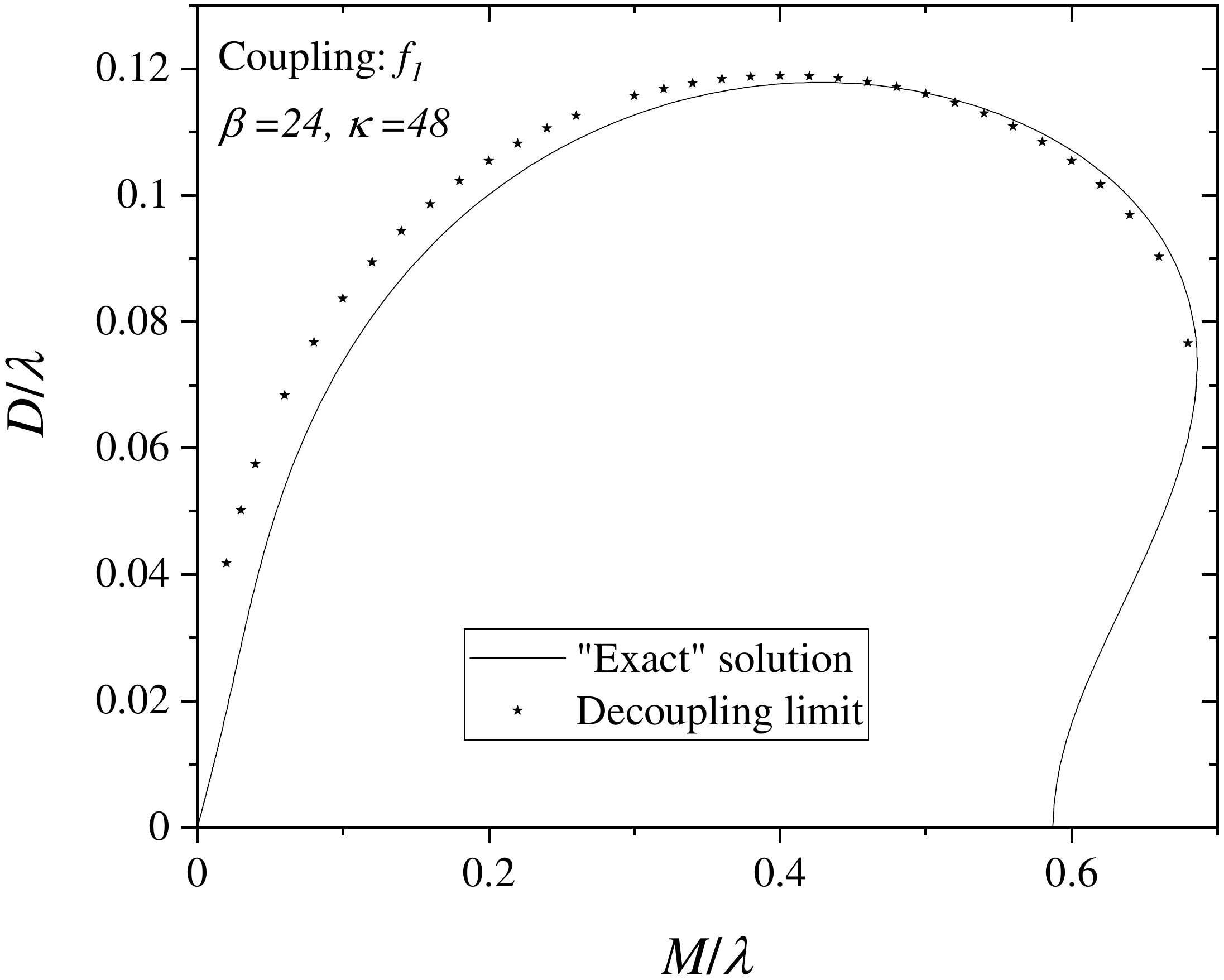}
\includegraphics[width=0.48\linewidth]{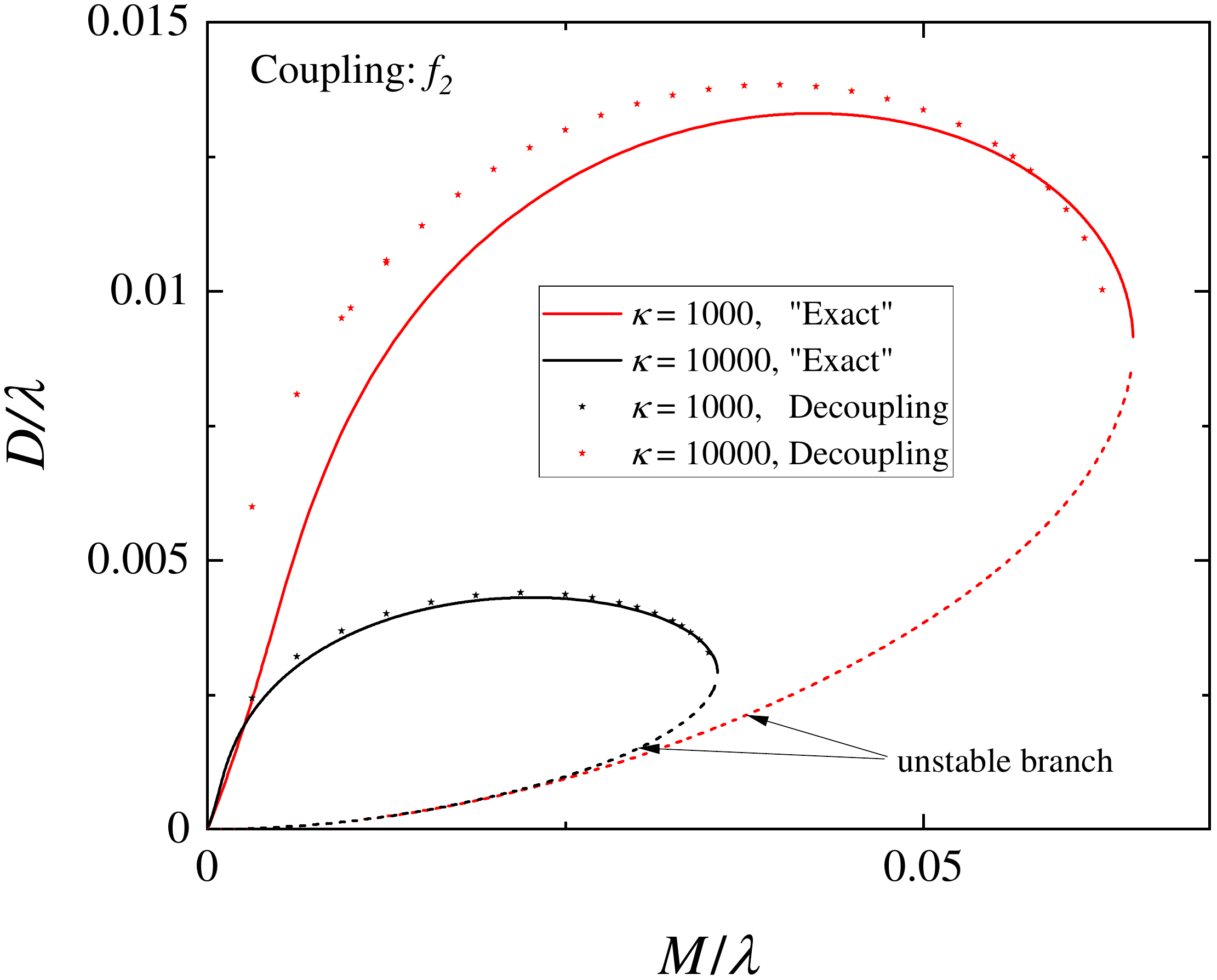}
\caption{Domain of existence obtained by evolving the scalar field on a Schwarzschild ($j=0$) background and by solving the static backreacting equations. \emph{Left:} Coupling $f_1$ with quadratic and quartic exponential terms. \emph{Right:} Coupling $f_2$ with only a quartic exponential term, for which no probe limit exists. }
\label{Fig:0}
\end{figure}

To investigate the evolution around a rotating black hole we proceed as the following: loop through increasing $j$ and increasing $M/\lambda$. The maximum value of the mass for a particular spin parameter is obtained once scalarization stops occurring. We run $j$ up to $j=0.995$, as $j=1.0$ corresponds to extremality and the closer we get to this value the more challenging it becomes to deal with the numerics.

The diagram $M/\lambda$ vs. $D/\lambda$ obtained with the evolution of the scalar field is shown on Fig. \ref{Fig:1} for the two couplings and several values of the spin parameter in the range $j\in[0,0.995]$. For both coupling functions, as far as only stable stationary configurations of the scalar field are concerned, there is no probe limit except for the trivialized Minkowski point, meaning $D/\lambda=0\iff M/\lambda=0$. Note that stationary configurations are found for the whole set of $j$ in this decoupling limit. At least for the backreacting solutions known to this day, this fails to be true because the condition for the regularity of the metric functions and the scalar field at the horizon is not satisfied for certain ranges of parameters \cite{PhysRevLett.123.011101,Collodel_2020,PhysRevLett.126.011104,PhysRevLett.126.011103}. If the backreaction of the scalar field on the metric is not taken into account, though, these conditions are always satisfied. Thus, even though the decoupling limit results are very useful and accurate for studying scalarized configurations, only the solution of the full system can demonstrate rigorously the existence domain for such scalarized black holes. This is a study underway.

Another particular trait we observe here is that, the higher the spin parameter, the larger the range in $M/\lambda$ is. This is in a way expected in the case of coupling $f_1$, because it was already proven in \cite{PhysRevLett.123.011101} that the bifurcation (at least in the case of $\kappa=0$) moves to larger  $M/\lambda$ as $j$ increases. We have explicitly checked that this is also true for $\kappa=48$, that we consider. Thus, not only the bifurcation point moves, but also the point of maximum  $M/\lambda$ of the upper stable scalarized branch. What is more interesting,  this behavior remains true also for the coupling $f_2$, the case of nonlinear scalarization in which Kerr is linearly stable in the full parameter range and no bifurcation point is present. The possibility of having scalarized black holes for much larger $M/\lambda$ in the near extremal case compared to the nonrotating BHs seems to be a general feature of scalarization, independent of whether it is a standard or a nonlinear one.

The scalar charge behaves nonmonotonically with respect to the background mass and presents a global maximum for every value of the spin parameter. As $j$ increases, the lower the value of $\max D/\lambda$ is and for lower values of the mass it occurs. Overall, for a fixed mass, the scalar charge gets smaller the faster the black hole rotates, but only up to a point. We observe that for $j\geq0.9$ there is always a point where the curves intercept and from that point on higher spin means higher charge for a fixed mass. Thus, there are regions of nonuniqueness. All of the properties just discussed above are common for the three cases analyzed as found in both panels of Fig. \ref{Fig:1}. Again, the main difference between the two couplings we promptly see is the range of $M/\lambda$ for which scalarized black holes exist. As commented, though, this is an artefact of the fact that we want to stay in a region of the parameter space where the scalar hair is weak enough so that the decoupling limit approximation is accurate enough.

\begin{figure}[]
\centering
\includegraphics[width=0.48\linewidth]{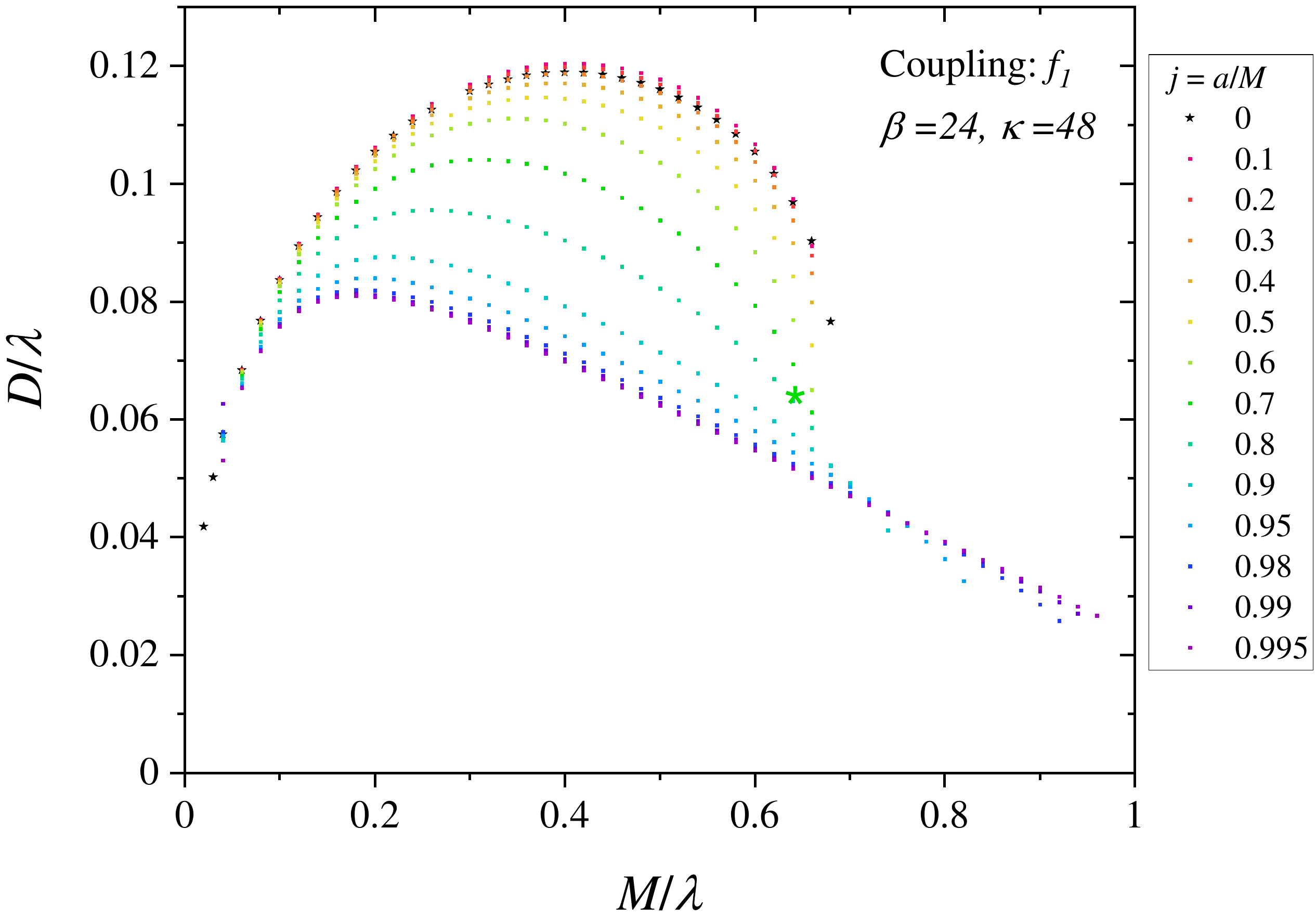}
\includegraphics[width=0.48\linewidth]{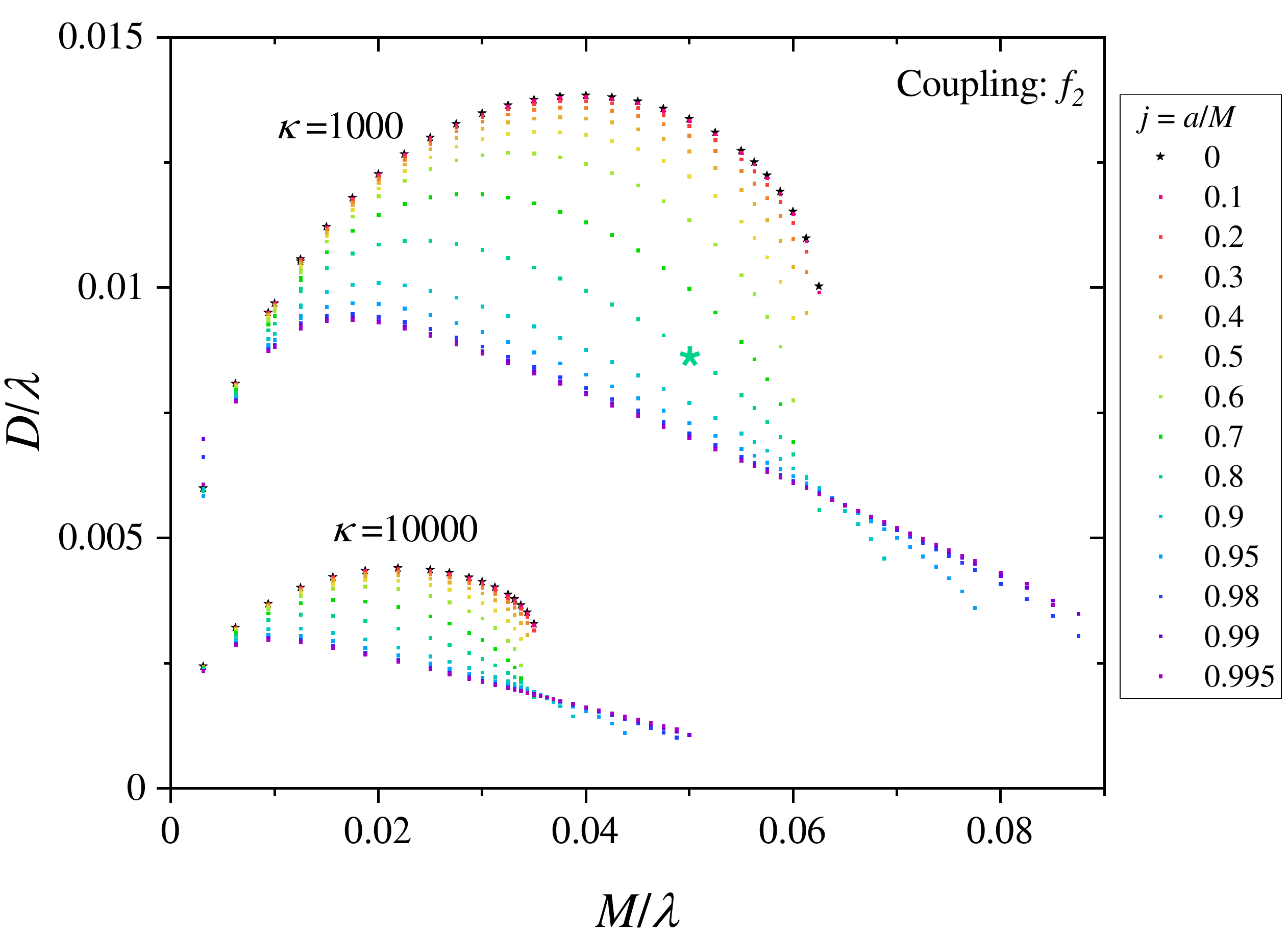}
\caption{Domain of existence obtained by evolving the scalar field on a Kerr background with several different spin parameters. \emph{Left:} Coupling $f_1$ with quadratic and quartic exponential terms. \emph{Right:} Coupling $f_2$ with only a quartic exponential term. The specific models for which we present the evolution later are marked with green stars. }
\label{Fig:1}
\end{figure}

Let us now turn to the scalarization process itself. As already discussed above and shown in Fig. \ref{Fig:0} for the Schwarzschild case, exact solutions of scalarized BHs form stable and unstable branches. For the range of parameters where we have only one stable scalarized solution and Kerr is linearly unstable (possible only for coupling $f_1$), the evolution is generally independent of the initial condition and arbitrary small perturbation will ignite the scalar field.  The timescale with which the saturation is achieved increases with decreasing initial amplitude, but the final state is always the same. However, if for the same parameter values that define the background an unstable scalarized solution exists (and Kerr is linearly stable), the final fate of the evolution becomes dependent on the initial configuration. In practice, if the starting amplitude is too small, then the Kerr black hole is stable against the nonlinear effects and the field decays. Else, if it meets its scalarization threshold, any initial amplitude higher than it will evolve to a nontrivial saturation and therefore a hairy black hole. In addition, the new hairly black holes are expected to be energetically more favourable in most of the parameter space compared to their GR counterpart \cite{PhysRevD.105.L041502}. This is exactly the mechanism of nonlinear scalarization observed first in the static case  that now translates also to rapid rotation.

In Fig. \ref{Fig:2} we present an example of the above mentioned nonlinear scalarization phenomenon for the theory defined by $f(\psi)=f_1(\psi)$. In the left panel the amplitude $\max\psi(t=0)=0.1$ is too small and the field decays. In the right, it is one and a half times higher $\max\psi(t=0)=0.15$ and scalarization is achieved. In this selected example the mass is given by $M/\lambda=0.63$ while $j=0.8$. When scalarized, the final charge amouts to $D/\lambda=0.065$. This solution is highlighted in Fig. \ref{Fig:1} (left panel). Each interior plot corresponds to a slice of $\varphi,t=$constant, where the $x,y$ axis correspond to $r\sin\theta/M,r\cos\theta/M$, respectively. Shown therein is the absolute value of the scalar field, $|\psi|$, in the same logscale for which the lowest value was set to $10^{-9}$. Apart from the amplitude, the initial condition is the same Gaussian pulse. In the left panel, the scalar interacts with the horizon and falls in, while the outgoing signal forms nodes at about $t=20M$ and travels away. As one can see, the scalar field decreases steadily in an exponential manner. From the right panel, on the other hand, the whole process takes place much more rapidly and reaches stationarity at about $t=80M$. Once the signal reaches the horizon and the outgoing wave has travelled away, at $t=30M$, the scalar field consistently increases. Had the initial amplitude been higher than the maximum stationary value of $\max\psi(t\rightarrow\infty)=0.175$, it would then decrease to this value. Due to the high spin, the scalar field near the event horizon is larger the closer it is to the equatorial plane.

\begin{figure}[!ht]
\centering
\includegraphics[width=0.48\linewidth]{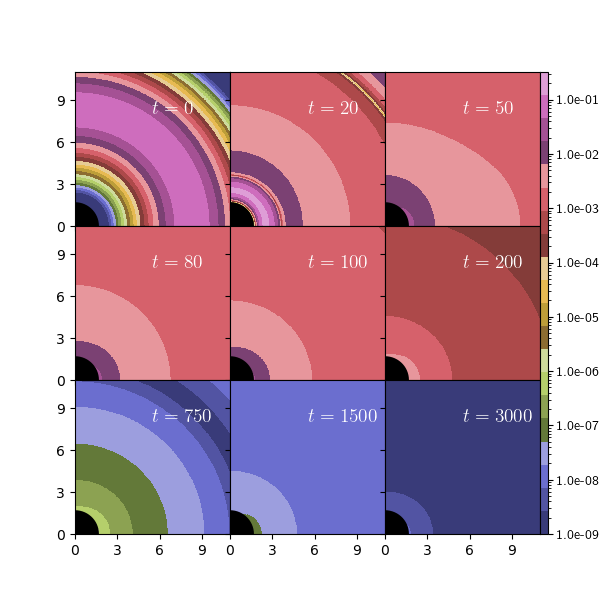}
\includegraphics[width=0.48\linewidth]{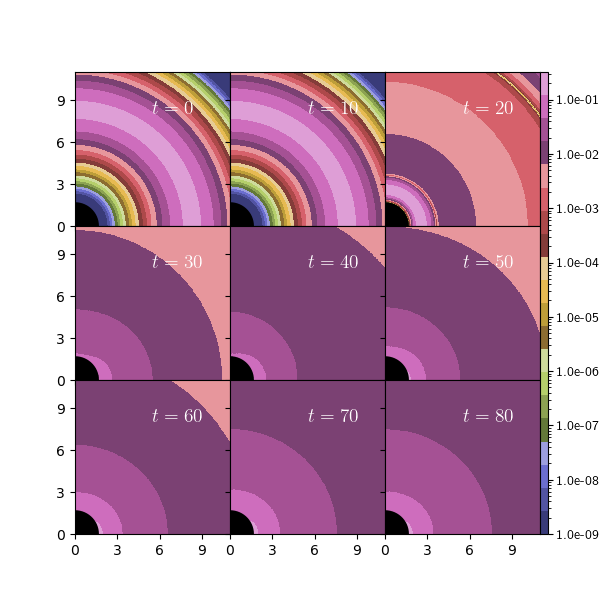}
\caption{ Different time steps in the scalarization process around a Kerr BH of $M/\lambda=0.63$ and $j=0.8$ at a constant $\varphi$ slice. This black hole model is marked with a star in the left panel of Fig. \ref{Fig:1}. The underlying theory is defined by $f(\psi)=f_1(\psi)$ and this is the model marked with green star in the left panel of Fig. \ref{Fig:1}. On the horizontal axis we have $r\sin\theta/M$, while on the vertical $r\cos\theta/M$. In colors is the magnitude of the scalar field $|\psi|$. The initial amplitudes of the Gaussian pulse are $\max\psi(t=0)=0.1$ (left) and $\max\psi(t=0)=0.15$ (right). The outcome of evolution for these parameters is dependent on the initial conditions. If the amplitude is not large enough, no scalarization occurs, and although the decaying process is long, it happens steadily.}
\label{Fig:2}
\end{figure}

Another example, this time for the pure quartic coupling $f_2$ with $f(\psi)=f_2(\psi)$ with $\kappa=1000.0$, that admits only the case of nonlinear scalarization and Kerr black hole is always linearly stable is given in Fig. \ref{Fig:4}. The axis and colorscheme are the same as above. 
In this Figure, we present two different BHs of fairly different normalized masses and charges, but with the same spin parameter $j=0.8$. On the left panel, the BH has $M/\lambda=2.1\times10^{-3}$ and scalarizes with a final $D/\lambda=2.4\times10^{-4}$. The solution lies very close to the lower left corner of Fig. \ref{Fig:1} (right panel), and is therefore not shown in there. The Gaussian pulse has an initial amplitude of $\max\psi(t=0)=0.01$. Smaller amplitudes can still reach the same stationary state, but at a much slower pace. Similar to the case discuss above there is a minimum amplitude below which the scalar hair can not develop. The configuration found at the end state differs greatly from the other cases in that the scalar field decays for much larger $r/M$ and the fact that for small values of $\theta$, the non-monotonic behavior of the scalar field with respect to the radial coordinate is quite prominent. As can be seen from the lower panels of the figure, near the vertical axis the scalar field first increases with increasing $r/M$. Such trait is not exclusive of this coupling and it is enhanced as the spin increases and the normalized mass decreases.
Note that in this case, the final state of the scalar field falls off much slower with $r$, but it is not constant as one might be misled by the lower panels of the figure. Instead, this is an artifact from the resolution. We chose to work with a single colorscheme and logarithmic scale for all the examples, aiming to achieve the most informative display possible. Nevertheless, it was not possible to do so without compromising the readability of some of the subplots. In the region where the scalar field is profiled by the color of maximum amplitude (in light pink), its value ranges within $0.21\leq|\psi|\leq0.3$.

On the right panel of Fig. \ref{Fig:4}, the BH's mass reads $M/\lambda=0.05$ and its charge $D/\lambda=0.0087$, while the initial amplitude of the pulse is of $\max\psi(t=0)=0.2$. This solution is highlighted in Fig. \ref{Fig:1} (right panel). In this case, similar to what was shown in Fig. \ref{Fig:2}, the stability of the background and the outcome of the evolution are sensitive to the initial condition. A smaller amplitude, e.g. $\max\psi(t=0)=0.1$, is incapable of scalarizing the BH and the scalar field decays completely with time.

\begin{figure}[!ht]
\centering
\includegraphics[width=0.48\linewidth]{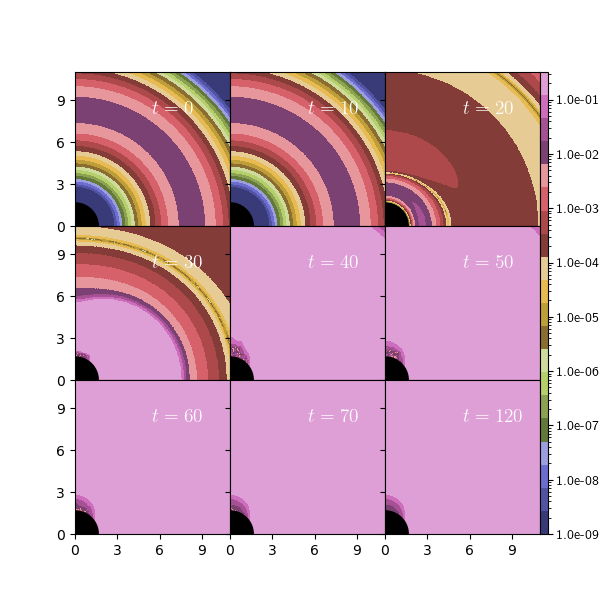}
\includegraphics[width=0.48\linewidth]{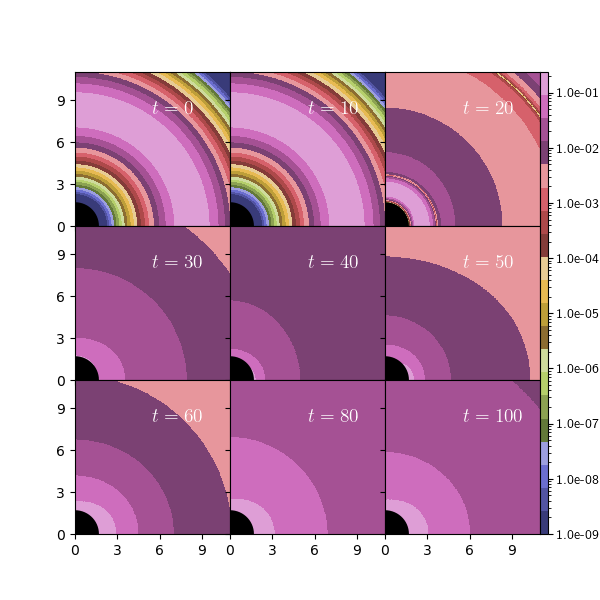}
\caption{Different time steps in the scalarization process around a Kerr BH within the context of the theory given by $f(\psi)=f_2(\psi)$. In both cases $j=0.8$. \emph{Left:} The BH has a normalized mass of $M/\lambda=0.0021$ and the initial amplitude is of $\max\psi(t=0)=0.01$. At the end state the scalar field falls off much more slowly than in the other cases depicted. \emph{Right:} The BH has a normalized mass of $M/\lambda=0.05$ and the initial amplitude is of $\max\psi(t=0)=0.2$. This black hole model is marked with a star in the right panel of Fig. \ref{Fig:1}. Just like the example presented before, the end state of the evolution on this background is sensitive to the initial conditions and no scalarization occurs for smaller amplitudes.}
\label{Fig:4}
\end{figure}

\section{Conclusions}
\label{sec:conclusions}

In this work we investigated the spontaneous nonlinear scalarization of Kerr BHs in EsGB theories. This was done by evolving in $2+1$ the scalar field with an initial configuration of a Gaussian pulse on a fixed background in repetitive form, sweeping through the parameter space. Due to the nonlinear nature of the equation, when the background is linearly unstable, instead of blowing up, the scalar field saturates approaching a stationary regime. Thanks to this, we can explore the whole range of parameters defining the BH and extract its scalar charge, as opposed to the linearized case. When no tachyonic instability is present, i.e. Kerr is always linearly stable, and for a properly chosen coupling between the scalar field and the Gauss-Bonnet invariant, such probe limit only occurs at zero curvature (Minkowski). There is therefore a gap between bald BHs and hairy solutions. The scalar field in this case can be excited only by a larger enough (nonlinear) scalar field perturbation that points towards the existence of nonlinear scalarization. In the present paper we aimed at generalizing this process to the case of rotating spacetimes.

Stationary scalar field configurations were found for the whole range of the spin parameter (up to $j=0.995$) that we could probe. We note, however, that since we neglect the backreaction of the scalar field on the spacetime metric, there is no regularity condition that the scalar field needs to satisfy. Therefore, it is to be expected that in the fully nonlinear coupled regime, the existence domain in the parameter space might be narrower, as in the case with pure tachyonic instabilities. On the other hand, though, we have been very careful to chose appropriate ranges of parameters and couplings, for which the scalar field is weak enough and thus the obtained approximate numerical solutions are quite close to the exact ones. This is something we explicitly demonstrate in the nonrotating case.

Our results show that the domain of existence of scalarized solutions quickly expands as the limit of extremal rotation is approached. Generally, the magnitude of the scalar charge decrease, though, with increasing the rotation. This qualitative behavior can change only for very rapid rotation (say $j>0.9$) and close to the maximum mass of the existence of the scalarized black holes, where rapid rotation can slightly enhance the scalar field.

For a coupling function allowing for linear destabilization of Kerr, and for most part of the parameter space, the final state of the evolution is independent of the initial conditions, which simply dictates the characteristic timescales at which saturation occurs. However, for couplings and parameters for which nonlinear scalarization is possible, the scalar field completely decays if the initial amplitude is below a certain value. In other words, in such case the Kerr background is linearly stable against scalar perturbations. As experience from spherically symmetric solutions shows (see e.g. right panel of Fig. \ref{Fig:0}), this hints the existence of two stationary hairy solutions of the same $M/\lambda$, one belonging to an unstable branch and the other to a stable one. This matter is currently under scrutiny and will appear in a future work on backreacting stationary solutions. What we could learn already from the decoupling limit results, though, is that the final scalar field profile can be qualitatively different for different  values of mass and spin. Rapidly rotating BHs of low mass portray a scalar field that behaves nonmonotonically  with $r$ near the poles, where its absolute value first increases as one moves away from the hole to only later fall off. Hence, for small polar angles, the maximum of the scalar field is not at the horizon. This is in contrast to slowly rotating massive BHs, for which the field decays monotonically with increasing $r$ for every $\theta$ value.  Even though this is not specific for the considered coupling only, we have demonstrated some models where the the effect of local scalar field maximum outside the black hole horizon is somewhat strong.

\section{Acknowledgments}
\label{sec:ack}
LC and DD acknowledge financial support via an Emmy Noether Research Group funded by the German Research Foundation (DFG) under grant no. DO 1771/1-1. SY would like to thank the University of Tuebingen for the financial support. The partial support by the Bulgarian NSF Grant KP-06-H28/7 is acknowledged. 

\bibliographystyle{ieeetr}
\bibliography{biblio}
\end{document}